\documentclass[aps,amssymb,twocolumn,superscriptaddress,nofootinbib]{revtex4}
\usepackage{setspace}
\usepackage{graphicx}
\usepackage{psfrag}

\newcommand{\brho}{\beta^\rho}
\newcommand{\bz}{\beta^z}
\newcommand{\sigmabar}{\bar{\sigma}}

\begin{document}
\title{Critical Collapse of a Complex Scalar Field with Angular Momentum}

\author{Matthew W. Choptuik}
\affiliation{CIAR Cosmology and Gravity Program \\
     Department of Physics and Astronomy,
     University of British Columbia,
     Vancouver BC, V6T 1Z1 Canada}
\author{Eric W. Hirschmann}
\affiliation{Department of Physics and Astronomy,
     Brigham Young University,
     Provo, UT 84604}
\author{Steven L. Liebling}
\affiliation{Southampton College, Long Island University,
     Southampton, NY 11968}
\author{Frans Pretorius}
\affiliation{Theoretical Astrophysics,
     California Institute of Technology,
     Pasadena, CA 91125}

\begin{abstract}
We report a new critical solution found at the threshold 
of axisymmetric gravitational collapse of a complex scalar field  with
angular momentum. To carry angular momentum
the scalar field cannot be axisymmetric; however, its azimuthal
dependence is defined so that the resulting stress energy tensor and
spacetime metric {\em are} axisymmetric. The critical solution found is 
non-spherical, discretely self-similar with an echoing exponent 
$\Delta=0.42\pm4\%$, and exhibits a scaling exponent $\gamma=0.11\pm10\%$ in near 
critical collapse. Our simulations
suggest that the solution is universal (within the imposed symmetry class),
modulo a family-dependent constant phase in the complex plane.
\end{abstract}

\maketitle

%\section{Introduction}\label{intro}

The main purpose of this work is to
study the effect of angular momentum in 
axisymmetric critical collapse of massless scalar fields. Critical collapse refers
to the {\em threshold} of black hole formation, where interesting
effects known as critical phenomena~\cite{choptuik} have
been observed in the gravitational collapse of a wide variety of 
types of matter, as well as vacuum gravitational waves (see \cite{creviews} for recent reviews). For 
spherically symmetric massless scalar collapse, 
this behavior includes
universality,
%\footnote{However, recent axisymmetric simulations\cite{paper2} 
%suggest that non-spherical features in the initial data may alter the
%nature of the critical solution},
scale invariance, and power law scaling of length scales that arise
near criticality.
In super-critical collapse, the characteristic length is
the mass, $M$, of black holes that form.
In the case of rotating collapse, since
angular momentum has dimension $\hbox{\em length}^2$, one
might naively expect the angular momentum, $J$, of the black holes
formed to scale as $J\propto M^2$. A more
refined analysis carried out using perturbation theory \cite{ggmg} 
suggests that
$J\propto M^{2\left(1-\mbox{Re}[\lambda']\right)}$,
where $\mbox{Re}[\lambda']$ is the real part of 
the exponent $\lambda'$ of the dominant perturbative mode that carries angular
momentum. In \cite{martin_garcia_gundlach}, $\mbox{Re}[\lambda']$ was found
to be roughly $-0.017$, implying an approximate scaling $J\propto M^{2.03}$.
Thus, the Kerr parameter $a=J/M^2$ is expected to scale to zero (albeit
slowly) as the black hole threshold is approached. Incidentally, this picture could be altered
significantly if the numerical evidence of \cite{paper2} proves correct,
and there is a {\em non-spherical} unstable mode of the critical
solution.

In general, numerical exploration of angular momentum in the collapse 
of a single real scalar 
field would require a 3D code,
for axisymmetric distributions of such a matter source cannot carry 
angular momentum. Constructing a general relativistic 3D simulation capable of 
resolving the range of length scales that unfold in scalar
field critical collapse is a daunting project, and may require computational
capacity not currently available. A ``cheaper'' alternative is to 
consider a set of distinct scalar fields, each with azimuthal 
dependence and hence angular momentum, and then add the different
fields coherently such that the net stress energy tensor is 
axisymmetric.
A natural way to achieve such a coherent sum is
via a single complex field, as will be explained in 
Sec.~\ref{system} (see \cite{UBC_group} for an alternative approach). 
One drawback to this method is that imposing such an ansatz for
the complex field forces a non-spherical energy distribution. 
This means
that the class of solutions we can study occupy a region
of phase space distinct from that of spherical spacetimes,
and so we cannot explore the role of angular momentum as
a perturbation in spherical critical collapse~\cite{choptuik,hirschmann_eardley}.
On a positive note, the fact
that we {\em do} find a new (axisymmetric) critical solution 
is interesting aside from questions of angular momentum, 
because it suggests that phase space has a more intricate structure
than one might have naively imagined, probably containing an infinite
set of distinct intermediate attractors characterized by their behavior 
near the center of symmetry (the results of~\cite{UBC_group} are also in
accord with this conjecture).
Regarding the question of how net angular momentum affects
threshold behavior in this model---it appears to be 
irrelevant, with the angular momentum of black holes formed
in super-critical collapse decaying significantly faster than
$J\propto M^2$, though our simulations are not accurate enough
to calculate exactly how fast.
Below we briefly describe the physical 
system and numerical code we use (for more details
see \cite{paper1,paper2,paper_mgamr,paper4}),
and present our results in Sec.~\ref{results}.

\vspace{-.1in}

\section{Physical System}\label{system}

We consider the Einstein field equations
\begin{equation}\label{einstein_tens}
R_{\mu\nu} - \frac{1}{2} R g_{\mu\nu} = 8\pi T_{\mu\nu},
\end{equation}
where $g_{\mu\nu}$ is the spacetime metric, $R_{\mu\nu}$ is the 
Ricci tensor, $R\equiv R^\mu{}_\mu$ is
the Ricci scalar, and we use geometric units with
Newton's constant $G$ and the speed of light $c$ set to 1.
We use a massless, minimally-coupled, complex scalar field $\Psi$ (with complex conjugate
$\bar{\Psi}$) as the matter source.
$\Psi$ satisfies a wave equation $\Psi_{;\mu}{}^{\mu} = 0$,
and has a stress energy tensor $T_{\mu\nu}$ given by
\begin{equation}\label{set_psic}
T_{\mu\nu} = \Psi_{;\mu}\bar{\Psi}_{;\nu} +
\bar{\Psi}_{;\mu}\Psi_{;\nu} -
g_{\mu\nu} \Psi_{;\gamma} \bar{\Psi}^{;\gamma}.
\end{equation}
We solve (\ref{einstein_tens}) and the wave equation (hereafter the field equations) 
in axisymmetry, using
coordinates $[t,\rho,z,\phi]$, where $\phi$ is adapted 
to the azimuthal symmetry, $t$ is time-like, and $(\rho,z)$ 
reduce to standard cylindrical coordinates in the flatspace limit.
The axial Killing vector is then
\begin{equation}\label{xi_def}
\xi^\nu=\left(\frac{\partial}{\partial\phi}\right)^\nu.
\end{equation}
The existence of this Killing vector allows us to define the
conserved angular momentum, $J$, of the spacetime 
\begin{equation}\label{l_tot}
J = -\int_\Sigma T_{\mu\nu} \xi^{\mu}n^{\nu}
\sqrt{h} \ d^3 x,
\end{equation}
where the integration is over the $t={\rm const.}$ spacelike hypersurface,
$\Sigma$, $h$ is the determinant of the intrinsic metric on $\Sigma$,
and $n^\mu$ is the hypersurface normal vector. Using (\ref{set_psic}) 
and (\ref{xi_def}), (\ref{l_tot}) evaluates to
\begin{equation}\label{J_exp}
J = -\int_\Sigma
\left[\Psi_{,\phi}\bar{\Psi}_{;\nu} +
\bar{\Psi}_{,\phi}\Psi_{;\nu} \right] n^{\nu}
\sqrt{h} \ d^3 x.
\end{equation}
Thus, for a configuration of the scalar field to have non-zero angular
momentum, $\Psi$ must have some azimuthal dependence. We thus adopt
the following ansatz
\begin{equation}\label{ansatz}
\Psi(\rho,z,t,\phi) \equiv \Phi(\rho,z,t) e^{im\phi}
\end{equation}
where $\Phi(\rho,z,t)$ is complex, and
$m$ must be an integer for the scalar field to be regular. It is 
straightforward to check that this form of $\Psi$ gives
a stress-energy tensor that is $\phi$-independent, yet
can yield net angular momentum.
Note that the on-axis ($\rho=0$) regularity condition
for $\Phi$ depends upon the value of $m$; specifically we must have
$\lim_{\rho\to0} \Phi(\rho,z,t) = \rho^m f(z,t)$.  For simplicity 
and specificity, we hereafter
restrict attention to the case $m=1$. As stated above, we expect 
that additional, distinct critical solutions exist for $m=2,3,\cdots$.
%\footnote{Specifically,
%$\Phi(\rho,z,t)$ must go to zero like $\rho^m$ in the limit $\rho\rightarrow 0$}

To keep our discussion concise, 
we only state the metric and set of variables we use (all functions of
$(\rho,z,t)$), and briefly describe the solution procedure---more details
can be found in \cite{paper1,paper2,paper_mgamr,paper4}. 
The line element is
\begin{eqnarray}
\label{metric}
ds^2= &-& \!\!\! \alpha^2 dt^2 \,\, + \\ \nonumber
      \psi^4 [ \left(d\rho + \brho dt \right)^2
      \!\! &+& \!\! \left(dz + \bz dt \right)^2 
      + \rho^2 e^{2\rho\sigmabar} d\phi^2 ]  \,\, + \\ \nonumber
      (\xi_\rho d\rho \!\! &+& \!\! \xi_z dz) 
      \left( 2 d\phi + \frac{\xi_\rho d\rho + \xi_z dz}{\psi^4 e^{2\rho\sigmabar} \rho^2}\right).
\end{eqnarray}
The lapse function, $\alpha$, is fixed  by maximal slicing, and~(\ref{metric}) 
reflects 
the additional coordinate conditions we have imposed: conformal
flatness of the two dimensional $\rho-z$ subspace, and 
$\xi_t=0$.
The Einstein equations
are written in first-order-in-time
form by introducing the following ``conjugate'' variables
\begin{equation}
\bar \Omega \equiv \left(- 2K_\rho{}^\rho - K_z{}^z\right)/\rho, \ \ \ \
\omega_\alpha \equiv \epsilon_{\alpha\beta\gamma\delta} \xi^{\beta} \xi^{\delta;\gamma},
\end{equation}
where $K_\alpha{}^\beta$ is the extrinsic curvature tensor
and $\omega_\alpha$ is 
the ``twist'' of the Killing vector. We separately evolve the real and imaginary
components of the scalar field by defining 
real functions $\Phi_r$ and $\Phi_i$ via
\begin{equation}\label{phi_def}
\Phi\equiv\rho(\Phi_r + i \Phi_i)
\end{equation}
and their dynamical conjugates $\Pi_r$ and $\Pi_i$ by
\begin{equation}
\Pi_r\equiv\mbox{Re}[\Phi_{,\alpha}n^{\alpha}]/\rho, \ \ \ \
\Pi_i\equiv\mbox{Im}[\Phi_{,\alpha}n^{\alpha}]/\rho.
\end{equation}
The factors of $\rho$ appearing in the above definitions
are included so that $\Phi_r$, $\Phi_i$, $\Pi_r$ and
$\Pi_i$ satisfy Neumann conditions on-axis. Similarly,
the variables corresponding to $\omega_\alpha$ and
$\xi_\alpha$ that are evolved in the code have appropriate
powers of $\rho$ factored out so that they 
satisfy Dirichlet conditions on-axis (see \cite{paper4} for
the specific definitions).

For the evolutions presented here we use the following initial data
for the scalar field components
\begin{eqnarray}
\Phi_{r|i}(\rho,z,0)&=&A_{r|i} 
 \exp\left[-\left(\sqrt{\rho^2+z^2}-R_{r|i}\right)^2/\Sigma^2\right] \, ,
\nonumber \\
\Pi_{r|i}(\rho,z,0)&=& \epsilon_{r|i} \Phi_{r|i}(\rho,z,0) \label{id} \, ,
\end{eqnarray}
where $A_r$,$A_i$,$R_r$,$R_i$,$\Sigma$,$\epsilon_r$ and $\epsilon_i$ 
are parameters fixing the shape of the initial scalar field profiles. 
All other freely specifiable variables are set to zero at $t=0$,
while the constrained variables $\alpha,\psi$ and $\lbrace\brho,\bz,w_\rho\rbrace$ 
are obtained by solving the maximal slicing condition, Hamiltonian
and momentum constraints respectively.

We use a partially constrained finite-difference scheme with adaptive
mesh refinement (AMR) to evolve the system of equations with time. In particular,
the slicing condition and momentum constraints are used to fix
$\alpha$ and $\lbrace\brho,\bz\rbrace$ respectively, while the remainder of the variables
are updated using their evolution equations.

\vspace{-.1in}

\section{Results}\label{results}

We now present results from a preliminary study of
the black hole threshold of the complex scalar field system
introduced in the previous section. We focus on four sets
of initial data, summarized in Table \ref{tab_id}.
\begin{table}
%\begin{center}
\begin{tabular}[t]{| c || c | c | c | c | c | c || c | c  |}
\hline
Label & $p$ & $R_r$ & $R_i$ & $\Sigma$ & $\epsilon_r$ & $\epsilon_i$
& $\delta_0$ & $\delta^\star$ \\
\hline
\hline
{\bf A}  & $A_r=A_i$   & 0.6  & 0.6 & 0.1 & 1 & -1 & $\pi/4$ & 0.91 $\pm$3\% \\
{\bf B}  & $A_r=A_i$   & 0.6  & 0.6 & 0.1 & 0 &  0 & $\pi/4$ & $\pi/4$ \\
{\bf C}  & $A_r=3A_i$  & 0.6  & 0.6 & 0.1 & 1 & -1 & $\tan^{-1}\frac{1}{3}$ & 0.39 $\pm$3\% \\
{\bf D}  & $A_r=A_i$   & 0.65 & 0.6 & 0.1 & 1 & -1 & --- & 1.34 $\pm$3\% \\
\hline
\end{tabular}
%\end{center}
\caption{
Parameters (see (\ref{id})) for the four families of
initial data discussed here, where 
$p$ denotes the parameter(s) we tune when
searching for the black hole threshold.
$\delta_0$ and $\delta^\star$ are the phase of the initial
data (if applicable) and estimated phase of the critical
solution respectively. All of the
simulations were run with the outer
boundary ($\rho_{\rm max},z_{\rm max},-z_{\rm min}$) at $4$
(except for the data of Fig.~\ref{lnjm_w_mf_h10}, where the outer boundary was 
at $16$), and in each case $p$ was tuned to within
$(p-p^\star)/p^\star \approx 10^{-14}$ of threshold.
}
\label{tab_id}
\end{table}
Family {\bf A} is the `canonical' example, consisting
of identical pulses of $\Phi_r(\rho,z,0)$ and $\Phi_i(\rho,z,0)$
that are initially approximately 
outgoing ($\epsilon_r=1$) and ingoing 
($\epsilon_i=-1$) respectively. This choice for
$(\epsilon_r,\epsilon_i)$ in a sense
maximizes the net angular momentum (\ref{J_exp}),
given the initial profiles for $\Phi_r$ and $\Phi_i$.
Conversely, family {\bf B} is time-symmetric, and hence
has zero net angular momentum. Families $\bf A$ and $\bf B$
can be written as $\Phi(\rho,z,0)=A(\rho,z) e^{i \delta_0}$,
where $A(\rho,z)$ is a real amplitude function and
$\delta_0$ is a {\em constant} phase factor, equal
to $\pi/4$ in both cases. Family {\bf C} is thus identical 
to family {\bf A} except for the initial phase.
For family {\bf D} $\Phi_r$ and $\Phi_i$ have
distinct initial profiles and thus cannot be characterized
by a constant phase. 

\begin{figure}
%\begin{center}
\includegraphics[width=5.0cm,clip=true,draft=false]{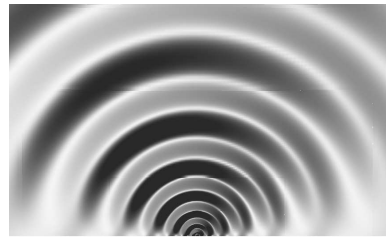}
%\end{center}
\caption
{A surface plot of the real part
($\Phi_r\rho$) of the complex field after several echoes
of a near-critical evolution.
The origin is at the bottom-center of the
figure, the $\rho$ axis runs vertically through the middle, and 
the $z$ axis runs horizontally. Only a single echo (roughly) at the origin
corresponds to the self-similar part of the spacetime---the
other ``waves'' visible were radiated during earlier echoes
of the field. Note also that the solution is {\em not} spherically
symmetric.
}
\label{w_m_h4_phi3_rho}
\end{figure}

\begin{figure}
%\begin{center}
%%BoundingBox: 18 164 572 690
\includegraphics[width=8.5cm,clip=true,draft=false]{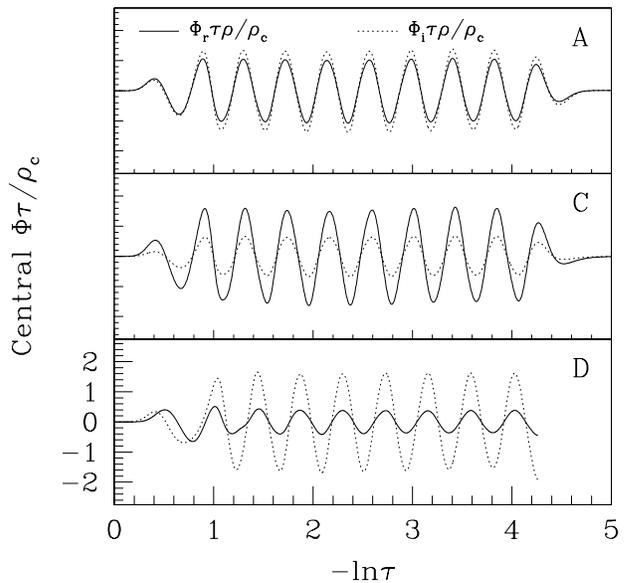}
%\end{center}
\caption
{The real ($\rho\Phi_r$) and imaginary ($\rho\Phi_r$) components of the 
central value of $\Phi$ (\ref{phi_def}) multiplied by proper time $\tau$
and divided by proper radius $\rho_c$ (\ref{rhoc}), versus 
$-\ln\tau$ for the near-critical collapse solutions of 
families {\bf A}, {\bf C} and {\bf D}
(the phase information for family {\bf B} is trivial as 
$\Phi_r=\Phi_i$ then, and so for brevity we do not
show it). The family {\bf D} solution shown here is super-critical,
and the simulation is stopped soon after an apparent horizon
is detected.
}
\label{phic}
\end{figure}

\begin{figure}
%\begin{center}
%%BoundingBox: 40 340 568 628
\includegraphics[width=7.5cm,clip=true,draft=false]{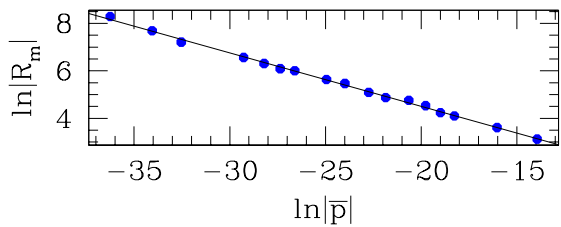}
%\end{center}
\caption
{$R_m$, the maximum value of the Ricci scalar on 
axis ($\rho=0$) attained during sub-critical evolution, versus 
distance $|\bar{p}|=p^\star-p$ from threshold,
for family {\bf A}.
Each point represents a single simulation. The
line is a least-squares fit to the data; $R_m$
has dimension $\hbox{\em length}^{-2}$, and hence the slope of the fit
is expected to be $-2\gamma$. For this case
we infer $\gamma\approx 0.11$. }
\label{lnpr_w_h4}
\end{figure}

Based upon the collapse simulations we have
performed for these four families of initial data, we
can {\em suggest} the following about the threshold
behavior for this matter model. There {\em is} apparently
a discretely self-similar critical solution that is 
universal to within a family-dependent phase. 
In other words,
one can write the critical solution $\Phi^\star$ for the 
scalar field as
$\Phi^\star(\rho,z,t)=A^\star(\rho,z,t)e^{i\delta^\star}$,
where $A^\star(\rho,z,t)$ is a universal 
real function and $\delta^\star$ is a family
dependent constant (see Table \ref{tab_id}). 
Note that this phase dependence is a consequence
of the $U(1)$ symmetry of the Lagrangian of the complex 
field, and has been observed in
charged scalar field critical collapse\cite{charged_cc}.
Also, note that {\em any} self-similar solution is unique only
up to a global rescaling of the form 
$(\bar{t},\bar{x}^i) \rightarrow (\kappa\bar{t},\kappa\bar{x}^i)$
when written in suitable coordinates $(\bar{t},\bar{x}^i)$,
with $\kappa$ a constant.
Fig.~\ref{w_m_h4_phi3_rho} shows
a snapshot of the real part of the scalar field ($\Phi_r\rho$) at
late times in a near-critical collapse simulation. 
%Note that
%the solution is {\em not} spherically symmetric.
To estimate $\delta^\star$ for a given family, and the echoing 
exponent $\Delta$ for the putative critical solution, we examine
the central value of the real and imaginary parts of the 
scalar field divided by proper radius $\rho_c$ (to factor
out the leading order approach to $0$ of $\Phi$ in 
a covariant manner) 
\begin{equation}\label{rhoc}
\rho_c\equiv\rho\psi^2 e^{\rho\sigmabar}.
\end{equation}
Fig.~\ref{phic} shows plots of $\Phi_r\tau\rho/\rho_c$ and
$\Phi_i\tau\rho/\rho_c$ vs. $-\ln\tau$ for the nearest-to-threshold 
solutions found, where $\tau$ is central proper
time (see \cite{paper2} for our definitions and 
computations of quantities measured by central
observers, but note that we define $\tau$
such that the accumulation event of the critical solution
corresponds to $\tau=0$).
We have multiplied the scalar field by $\tau$
in these plots to cancel the artificial growth
introduced by dividing by $\rho_c$.
Note that the equations of motion for $\Phi_r$ and $\Phi_i$
are identical; hence if $\Phi_r(\rho,z,0) = \Phi_i(\rho,z,0)$ 
(as with family {\bf B}), then
the initial phase, $\delta_0=\pi/4$, is preserved during evolution.
The echoing exponent $\Delta$ is the period of the self-similar
solution in logarithmic proper time;
from Fig.~\ref{phic} (and similar data for family {\bf B}) we
estimate $\Delta=0.42\pm4\%$. 
To estimate the scaling exponent $\gamma$, we measure how the
maximum value attained by the Ricci scalar (on axis), $R_m$, in
sub-critical evolutions 
depends upon the parameter-space distance from threshold, 
$p^\star - p$~\cite{garfinkle_duncan}. 
Representative results are shown in Fig.~\ref{lnpr_w_h4}. Combining such
data from all the families, we estimate $\gamma=0.11\pm 10\%$. 
For a discretely self-similar solution, one expects the linear
relationship assumed in Fig.~\ref{lnpr_w_h4} to be modulated
by an oscillation of period $2\Delta$ \cite{wiggle}; we have not run a sufficient
number of simulations to adequately resolve such an oscillation.

The uncertainties quoted above for $\gamma$, $\Delta$ 
and $\delta^\star$ (in Table \ref{tab_id})
were estimated from convergence calculations from simulations
using 3 different values of the maximum truncation error threshold 
that controls the AMR algorithm, 
but do not account for possible systematic errors (see a discussion
of related issues in \cite{paper2}).

\begin{figure}[b]
%\begin{center}
%%BoundingBox: 40 160 612 628
\includegraphics[width=7.5cm,clip=true,draft=false]{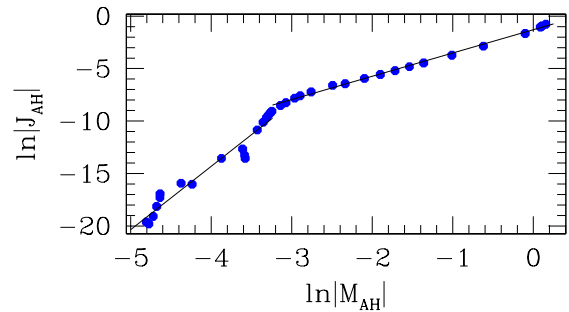}
%\end{center}
\caption
{Estimated black hole mass ($M_{{\rm AH}}$) versus angular momentum
($J_{{\rm AH}}$) in super-critical collapse of family {\bf A} initial data.
Points represent individual simulations, while
the two lines are separate linear regression fits to the set of points
to the left and right of the ``knee'' in the curve at $\ln(M_{{\rm AH}})\approx -3.2$,
with slopes $\approx 6.0$ and $\approx 2.2$ respectively. In $\ln \bar{p}$,
the horizontal scale ranges from $-22$ to $-2$ (compare to Fig.~\ref{lnpr_w_h4}).}
\label{lnjm_w_mf_h10}
\end{figure}

Finally, as mentioned in the introduction, for the near-critical
solutions described here, 
net angular momentum seems to be completely irrelevant. To within the accuracy
of our simulations, we cannot differentiate between the late
time, self-similar regions of the spacetimes obtained from families
{\bf A} or {\bf B}, and in the latter case, there is {\em no} 
angular momentum.
In fact, any angular momentum present
is radiated away so rapidly 
during self-similar collapse that we cannot accurately 
calculate the corresponding scaling exponent (i.e. the
remaining angular momentum is zero to within truncation error). Fig.~\ref{lnjm_w_mf_h10}
shows a plot of the mass estimate $M_{{\rm AH}}$ versus angular 
momentum $J_{{\rm AH}}$, on a logarithmic scale, of black holes 
formed in super-critical collapse.
$M_{{\rm AH}}$ and $J_{{\rm AH}}$ are calculated from the area
and angular momentum of the apparent horizon respectively (using 
the dynamical horizon framework \cite{dynamical_horizon}), and are computed
at the time the apparent horizon is {\em first} detected. 
For large black holes (i.e. those with $M_{\rm AH}$ of order the ADM mass),
Fig.~\ref{lnjm_w_mf_h10} suggests that
$J_{{\rm AH}} \propto M_{{\rm AH}}{}^2$. However, this region of parameter space
is ``maximally'' far from threshold, in that these are almost the largest
black holes that we can form from initial data not already containing an
apparent horizon.
For somewhat smaller black holes, Fig.~\ref{lnjm_w_mf_h10} shows
a transition to a relationship closer to $J_{{\rm AH}} \propto M_{{\rm AH}}{}^6$;
however, we are still far from threshold there, and furthermore are
entering the regime where the angular momentum calculation is dominated by
numerical errors. Thus, perhaps the only quantitative statement we
can make regarding angular momentum scaling for this system
is that $J$ scales to zero significantly faster that $J\propto M^2$.

\smallskip

\section*{Acknowledgments}

The authors gratefully acknowledge research support from 
CIAR, NSERC,  NSF PHY-9900644, NSF PHY-0099568, NSF PHY-0139782, 
NSF PHY-0139980, Southampton College and 
Caltech's Richard Chase Tolman Fund. Part of this work was
completed at the KITP, supported by NSF PHY99-07949.
Simulations were performed on UBC's {\bf vn}
cluster, (supported by CFI and BCKDF), and the {\bf MACI} cluster
at the University of Calgary (supported by CFI and ASRA).

\end{document}